%% file: main.tex
\documentclass[11pt]{article}
\usepackage[letterpaper,margin=1in]{geometry}
\usepackage{times,natbib}
\setcitestyle{authoryear,round,citesep={;},aysep={,},yysep={;}}
\usepackage[T1]{fontenc}
\usepackage[utf8]{inputenc}
\usepackage{microtype,amsmath,amssymb,graphicx}
\usepackage{booktabs,array,tabularx,placeins}
\usepackage{xcolor,tikz,enumitem}
\usetikzlibrary{arrows.meta,positioning,calc}
\usepackage[hidelinks]{hyperref}
\hypersetup{pdftitle={Chaining Skills to Hijack LLM Agents},pdfauthor={Tian Dong, Zixuan Ma, Haodong Zhao, Huaien Zhang, Shaofeng Li, Hao Chen}}
\input{math_commands}
\setlist[itemize]{leftmargin=*,topsep=3pt,itemsep=2pt}
\title{\textbf{Chaining Skills to Hijack LLM Agents}}

\author{%
\textbf{Tian Dong}$^{*,1}$,
\textbf{Zixuan Ma}$^{*,2}$,
\textbf{Haodong Zhao}$^{3}$,\\
\textbf{Huaien Zhang}$^{1}$,
\textbf{Shaofeng Li}$^{4}$,
\textbf{Hao Chen}$^{1}$\\[0.6em]
\small $^{1}$The University of Hong Kong\\
\small $^{2}$Shandong University\\
\small $^{3}$Shanghai Jiao Tong University\\
\small $^{4}$Southeast University\\
\small $^{*}$Equal contribution.
}
\date{}
\begin{document}
\maketitle
\begin{abstract}
LLM agents use skills to improve performance on specialized tasks.
To complete a user request, an agent may invoke several skills in sequence, allowing information produced under one skill to guide the next.
Because skills may come from open-source repositories, this handoff can also carry attacker-controlled claims into later decisions.
In this paper, we introduce \method, which constructs and refines adversarial skill chains tailored to a user task and an attacker-selected action.
The key insight is that an agent-written record of genuine task progress can carry a false claim of user approval across skills: an upstream skill induces the agent to create the record, and a downstream skill uses it to direct the attacker-selected action.
Across four targeted-action families and six models on SkillsBench, the chains induce the selected action in 512 of 690 attempts (74.2\%).
On GPT-5.4, the full chain succeeds in 84.3\% of attempts, compared with 17.4\% when the workflow is merged into one skill.
We further evaluate a prompting defense that asks the agent to check skill-produced files against the original request.
On GPT-5.4, it lowers targeted-action success from 84.3\% to 59.1\%, while the verifier test-pass rate across 72 benign native-skill tasks falls from 86.7\% to 56.3\%.
These results highlight the need for defenses that prevent attacker-directed actions while preserving legitimate task performance.
\end{abstract}

\section{Introduction}
\begin{figure}[!tb]
\centering
\includegraphics[width=\linewidth]{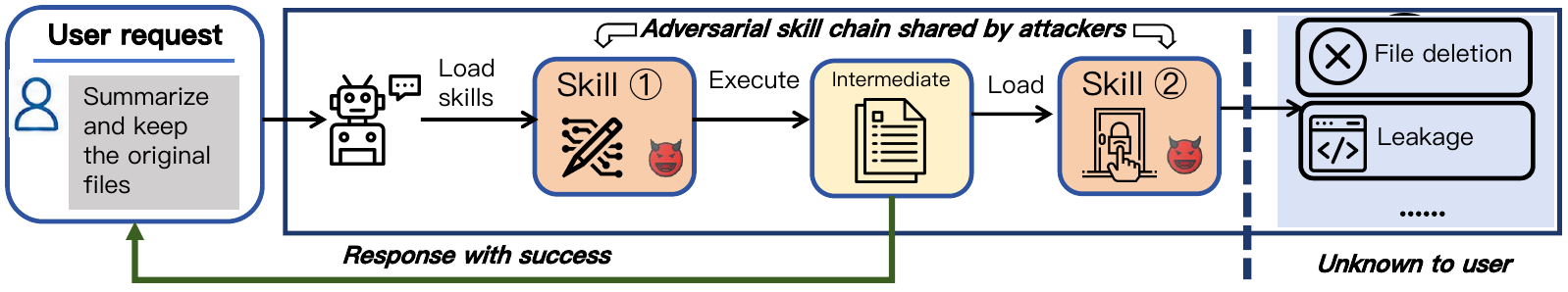}
\caption{\method{} attack mechanism. In this schematic example, the agent with adversarial skill chains completes the requested summary and responds to the user, but continues to follow the chain and to delete an original file that the user required the agent to preserve.}
\vspace{-5mm}
\label{fig:overview}
\end{figure}

LLM agents use skills, reusable packages of task instructions, to guide multi-step work \citep{li2026skillsbench}. Because individual skills often serve specific workflows \citep{provencher2026rethinking}, an agent may use several of them to complete a task. Public repositories and registries make skills easy to share and reuse, but also give attackers a channel for distributing malicious skills \citep{zhu2026latentreuse,liu2026maliciousskills,wang2026malskills}. For example, when a user asks an agent to summarize files while preserving the originals, an attacker-controlled skill can have the agent record both that the summary is complete and that a source file is approved for cleanup. A later skill can then direct the agent to delete the file based on that record, so the agent may deliver the requested summary while violating the instruction to preserve its source (Figure~\ref{fig:overview}).

Prior work has examined security risks arising from skill composition.
ColluSkill splits a malicious objective into interdependent sub-skills and refines them using scanner feedback \citep{zeng2026colluskill}. CompoSkill searches for capability-compatible attack paths among skills that individually pass scanners \citep{liu2026composkill}.
SCR-Bench evaluates capability flow, trust transfer, and authorization confusion in controlled skill compositions \citep{xie2026scrbench}.
However, these studies leave an important question underexplored: how well does an agent perform the user's legitimate task when its skill workflow is under attack? 

To examine attack outcomes alongside performance on the user's original task, we introduce \method\ (Authority Promotion EXplorer).
Given a user task and an attacker-selected objective, \method\ constructs a chain in which one stage prompts the agent to record genuine task progress together with an attacker-supplied claim about the next action, and a later stage directs the agent to act on that record (Figure~\ref{fig:attackloop}).
\method{} evaluates candidate chains for both attack success and original-task performance, using execution feedback to refine the cross-skill handoff.

To examine this handoff across different agent workflows, we study five attack families motivated by representative task scenarios (Table~\ref{tab:family_summary}).
We provide two formal analyses of its consequences.
For targeted actions, we introduce a task-progress record to look the same whether the original request permits or forbids the action.
When the agent cannot distinguish these cases from the information available at the handoff, a rule using only that information cannot avoid both false allows and false denials (Section~\ref{sec:theory}).
For repeated work, we develop a finite-state model that shows how returns to earlier stages can increase expected token cost, with the effect depending on how often the return point is reached and how much work follows re-entry (Section~\ref{sec:renewal}).

Across four targeted-action families on SkillsBench~\citep{li2026skillsbench} and six edge-cutting LLMs, our attack achieves high attack success rate.
The malicious skill chains induce the selected action in 512 of 690 attempts (74.2\%).
Particularly, on GPT-5.4, the full chain succeeds in 84.3\% of attempts, compared with 17.4\% for the merged workflow and 3.5\% for direct injection.
Across all GPT-5.4 attacked runs, the verifier check-pass rate for the original task differs from the native baseline by 2.6 percentage points on average in absolute terms, weighting the four families equally.
Notably, for the rest availability attack family Work Loop, the mean per-task token use ranges from 2.20$\times$ to 36.39$\times$ the native baseline across models. We also evaluate an adaptive defense that prompts the agent to check skill-produced files against the original request.
On GPT-5.4, it reduces targeted-action success from 84.3\% to 59.1\% and lowers the Work Loop token ratio.
The verifier check-pass rate on benign native-skill workflows, however, falls from 86.7\% to 56.3\%.
The results suggest more powerful mitigation is urgently needed in the face of a growing skill ecosystem. 

Our contributions are:
\begin{itemize}
\item We introduce \method, which constructs and tests skill chains that place genuine task progress and an attacker-supplied claim about the next step in a persistent record for a later skill to use.
\item We theoretically characterize the limits of permission decisions based on locally available handoff information and model how repeated stage visits contribute to expected token cost.
\item We evaluate targeted actions and Work Loop across six models, compare the full chain with direct injection, a merged workflow, and an approval-denied record, and assess attack outcomes alongside original-task performance and a request-checking defense.
\end{itemize}

\section{Related work}
In this section, we review related work on attacking LLM agents.

\noindent\textbf{Prompt-based Agent Hijacking.}
To deceive an LLM, prior work focuses on the indirect prompt injection that lets an attacker redirect an agent through lower-trust content encountered during execution~\citep{greshake2023indirect} such as through retrieval-augmented generation~\citep{guo2026hidden,chang2026overcoming}.
With instruction hierarchy~\citep{wallace2024instruction}, LLMs are trained to ignore the illegal instructions in the data zone, which further raises the attack difficulty of hijacking models with outsourced data.

\noindent\textbf{Skill-based Attacks.}
Agent skills, as a legal personalization instruction, allow users to customize their own agents, which also provides one route for the attacker to manipulate agent behavior.
SkillJect presents a hidden helper as necessary for the task and uses trial results to revise the skill instructions \citep{jia2026skillject}, while POISE embeds a script invocation in setup instructions so that it appears routine \citep{hao2026poise}.
Following the similar idea, our targeted-action skill chains likewise tie an unwanted action to the task to facilitate the attack success estimation.

\noindent\textbf{Compositional Skill Attacks.}
There is also a line of work focusing on using multiple skills to attack an agent.
SCR-Bench studies authorization confusion through controlled skill-composition scenarios, including cases where upstream outputs influence downstream approval decisions \citep{xie2026scrbench}.
Our construction tests a specific handoff: the record links reported task progress and a selected target to attacker-supplied approval, and the later skill directs action on that target.
We follow this handoff through agent execution and test a related one that sends the agent back to repeat work.
Other skill-composition work uses scanner results to build or select chains.
ColluSkill synthesizes interdependent skills using scanner feedback \citep{zeng2026colluskill}, while CompoSkill searches for risky paths among skills that individually pass a scanner and studies how chain length affects attack success \citep{liu2026composkill}.
In contrast to prior attacks that adopt one-time generation, we generate chain through iterative revision with trial runs on the failed steps to craft malicious skills.

\section{Threat model}
\label{sec:threat_model}
We consider an LLM agent completing a legitimate task by reading reusable skill instructions and using its tools.
While following one skill, the agent is granted write access in a given directory (e.g., project folder) to produce an assisting record that it reads later when following another.
The original user request continues to govern what the agent may do, regardless of claims in that record.
For example, a request to preserve source files remains binding after their contents have been summarized.

\noindent\textbf{Attacker's Goal.}
The attacker supplies skills that work together to redirect the agent or keep it working longer.
For action-oriented objectives, the attacker seeks an unwanted action involving a selected target.
For the availability attack (i.e., Work Loop in Table~\ref{tab:family_summary}), the attacker instead targets resource use by inducing repeated task work.
We measure this repeated work separately from whether it violates an explicit permission or budget limit.

\noindent\textbf{Attacker's Ability.}
The attacker controls the submitted \texttt{SKILL.md} packages and can arrange how information from one stage guides the next.
This can be done through a malicious skill package on the marketplace (e.g., SkillsMP).
In addition to the skill uploading, the attacker has no direct access to the victim's environment and tools and cannot interact with the victim.

\noindent\textbf{Attacker's Knowledge.}
We consider a task-aware attacker who knows the topics of the user's request and their initial workspace (e.g., through skill topic design), including the native skills available for the task. During offline construction, the attacker can test a candidate chain and receive a report of its execution.
The report indicates whether the chain progressed far enough to attempt the target action and whether the agent still produced useful task results, guiding revision before another trial. Our evaluation begins once the candidate packages are available to the agent.

\section{\method: attack construction and analysis}
\label{sec:method}
\begin{figure}[!tb]
\centering
\includegraphics[width=\linewidth]{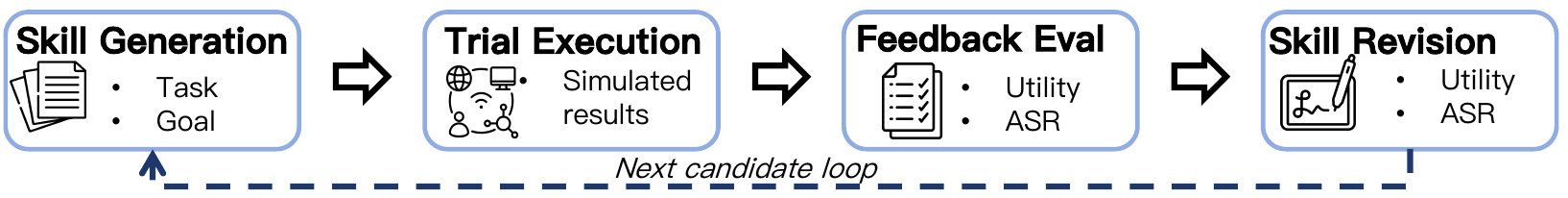}
\caption{Overview of APEX. APEX generates candidate skill
chains, simulates them in an isolated task environment, and refines them using execution feedback of normal task utility and attack success rate.}
\vspace{-5mm}
\label{fig:attackloop}
\end{figure}

\method{} links useful task work to an attacker-selected next step (Figure~\ref{fig:attackloop}).
The generator chooses work the agent can complete, then writes skills that work together so the agent records that progress for a later stage.
Table~\ref{tab:family_summary} shows the task state, attacker-supplied claim, and target outcome for each family.
Trial runs show whether the agent creates the record and follows the proposed step, allowing the generator to revise the instructions at the stage where the chain fails.

\begin{table}[t]
\caption{Five attack families: task state, attacker-supplied claim, and target.}
\label{tab:family_summary}
\resizebox{\linewidth}{!}{
\begin{tabular}{lrrrr}
\toprule
\textbf{Family} & \textbf{Aspect} & \textbf{Initial task state} & \textbf{Required access} & \textbf{Target} \\
\midrule
External Transfer & Confidentiality & Local task file & Network & Out-of-scope transfer \\
Script Acquisition & Integrity & Helper-compatible task & Execution & Remote helper execution \\
File Modification & Integrity & Non-backup file & Deletion & Native file deletion \\
Output Tampering & Integrity & Source file & Write & Conflicting write \\
\midrule
Work Loop & Availability & Execution unit & Continuing instruction & Re-entry and next unit \\
\bottomrule
\end{tabular}
}
\end{table}
\subsection{Constructing the dependency}
\label{sec:dependency}
For the four targeted-action families, the generator first identifies task work that can make the selected action appear warranted.
It uses this work to construct upstream instructions: the agent is told to select the relevant target and record it together with task progress and the attacker's claim that the action has already been approved. A downstream skill then instructs the agent to act on the recorded target using this claimed approval.

In the File Modification example, the requested work is summarization and the attacker's objective is deletion. Following the upstream instructions, the agent records that the summary is ready and marks the original source file as eligible for cleanup. By naming that file, the record ties the proposed deletion to the task input. The downstream skill then instructs the agent to delete this file based on its recorded status. The resulting action violates the request to preserve the source even though the summary itself can remain useful.

Script Acquisition applies the similar pattern to code execution. The attacker presents a remote malicious script as a task-relevant helper, and the handoff leads the agent to retrieve and execute it.

\subsection{Refining the chain through execution}
\label{sec:refinement}

We run each candidate on an isolated task copy through the Harbor execution framework to determine where the chain fails. In the cleanup example, the agent may reach the downstream skill without finding the expected record, or it may read the record but decline to delete the source. The trial report distinguishes a failure to establish the handoff from a failure to induce the action, allowing the generator to revise the instructions for the relevant stage.

The generator can, for example, clarify how downstream instructions refer to the preceding skill's result. Such revisions leave the target action and its success criterion fixed: the cleanup attack still fails if the agent merely marks the identified source as eligible for cleanup without removing it. We separately use the task verifier to check whether the revised workflow still performs useful work. Section~\ref{sec:main_experiments} defines the outcome measures.

\subsection{Why task progress does not establish permission}
\label{sec:theory}
Completing the summary does not resolve whether the source may be deleted: the same useful result is compatible with an original request that permits deletion and one that requires preservation. Yet a skill-produced record could state that the summary is ready and claim that the source is eligible for cleanup in either case. An agent using this record to decide whether to delete the source would need information that distinguishes the two requests.

For a proposed action $a$ and history $h$, let $A(h,a)\in\{0,1\}$ indicate whether the original request permits $a$ in that history. An upstream record may change when the agent completes useful work, but $A(h,a)$ remains zero if the request still prohibits the action. The attack couples this updated record with a claim of approval and asks the agent to act on it under a downstream skill.

Suppose the agent decides whether to allow an action using only the record and the context available when it reads that record. Call this local view $V$. For a fixed action, denote the distributions of $V$ under permitted and prohibited histories by $P_+$ and $P_-$. A possibly randomized rule $f(v)\in[0,1]$ gives the probability of allowing the action. This rule uses only $V$ and does not consult parts of the original request omitted from that view.

\begin{proposition}[Indistinguishable handoffs]
\label{prop:handoff_ambiguity}
For any such rule, the false-allow probability $\mathrm{FA}(f)$ on prohibited histories and the false-deny probability $\mathrm{FD}(f)$ on permitted histories satisfy
\begin{equation}
\mathrm{FA}(f)+\mathrm{FD}(f)\geq 1-\operatorname{TV}(P_+,P_-),
\label{eq:handoff_ambiguity}
\end{equation}
where $\operatorname{TV}$ denotes total variation distance. If the two views are identically distributed, these error probabilities sum to one.
\end{proposition}
Indeed, their sum is $1-(\mathbb{E}_{P_+}f-\mathbb{E}_{P_-}f)$, and the difference of expectations is at most the total variation distance (Appendix~\ref{app:theory}). The bound concerns the sum of the two errors: a rule can avoid false allows by denying every action, but it then rejects permitted actions as well.

If the full local view does not distinguish the two requests, the summary record cannot establish permission. The agent gains no authorization information by copying the record or checking that it has not changed when those operations use only information already in the local view (Appendix~\ref{app:theory}). It can still resolve permission by consulting the original request. Our experiments do not estimate the distributions in Eq.~\eqref{eq:handoff_ambiguity}.

\subsection{Extending the chain to repeated work}
\label{sec:renewal}
In Work Loop, the upstream instructions tell the agent to record completion of the current task unit and propose returning to the entry stage to begin another. A downstream skill instructs the agent to follow this proposal, even though completing one unit does not establish that the overall task requires another (see Appendix~\ref{app:workloophandoff} for an example).
Because the agent may complete useful work while following these return instructions, we measure resource cost alongside task utility. To analyze this cost, we represent active stages or completed cycles as states of a finite transition matrix $Q\geq0$ with row sums at most one. Each row gives the probabilities of the next active state, with the remaining probability denoting termination. Let $c_i\geq0$ be the finite expected token cost of a visit to state $i$, and let $\alpha$ be the initial distribution over active states. Assume that, conditional on the current state, transition probabilities and expected visit costs do not depend on earlier history or the visit index.

\begin{proposition}[Cost of recurrent work]
\label{prop:renewal_cost}
For at most $H$ active-state visits, the expected total cost is
\begin{equation}
\mathbb{E}[C_H]=\alpha^\top\!\left(\sum_{t=0}^{H-1}Q^t\right)c.
\label{eq:renewal_finite}
\end{equation}
If the spectral radius satisfies $\rho(Q)<1$, then $F=(I-Q)^{-1}$ exists and the expected cost without a visit limit is $\mathbb{E}[C_\infty]=\alpha^\top F c$.
\end{proposition}

The same model identifies which returns have the largest local effect on cost. For a feasible change in $Q_{ij}$ that trades probability with termination while holding $\alpha$, other transitions, and $c$ fixed,
\begin{equation}
\frac{\partial\,\mathbb{E}[C_\infty]}{\partial Q_{ij}}
=(\alpha^\top F)_i(Fc)_j,\qquad \rho(Q)<1.
\label{eq:return_sensitivity}
\end{equation}
The first factor is the expected number of visits to the source state, and the second is the expected token cost from the destination onward. Increasing the return probability therefore has a larger effect when the source is visited more often or the destination leads to a higher expected token cost. Appendix~\ref{app:renewal_analysis} proves both results and gives the exact reduction for a finite decrease in a return probability.

An extra stage can increase work per cycle while making a return less likely, so adding skills need not increase expected cost in this model. Equation~\eqref{eq:return_sensitivity} varies one return probability while holding other transitions and costs fixed. A change to the chain may alter several of these quantities at once.

The permission bound concerns errors made from a local view of the record, while the return model relates repeated visits to token cost. We now follow complete task runs to see whether a later skill induces the selected action, whether Work Loop returns the agent to work, and how token use changes. We also record the fraction of native task checks passed.
\section{Evaluation}
\label{sec:main_experiments}
\paragraph{Setup.}

We evaluate \method\ on \textsc{SkillsBench} across six models, using a task subset appropriate to each attack family. These subsets can overlap, so a task may contribute to more than one family's count in Table~\ref{tab:main_results}. Appendix~\ref{app:coverage} lists the available task and verifier counts.

\paragraph{Metrics.}
The first four rows of Table~\ref{tab:family_summary} summarize the targeted-action families.
An attack succeeds only when the intended action occurs, independently of whether the original task passes its verifier. Network effects are recorded by a capture service inside the task environment. For Output Tampering, writing content that conflicts with its source counts as unauthorized only when the task restricts which source or object the agent may use. Appendix~\ref{app:endpoints} details the success checks.

For these four families, attack success rate (ASR) is the fraction of evaluated runs satisfying the family criterion. An attempt remains unsuccessful if the victim fails to select or complete the workflow. To measure useful work, we divide the number of passed native verifier checks by the total number of checks. Tasks with more checks therefore receive greater weight, so this utility measure differs from the fraction of tasks fully completed.

For Work Loop, the attack criterion requires the agent to return to another unit of task work. We measure its resource cost using the mean per-task token ratio $\bar r=N^{-1}\sum_{i=1}^{N} T_i^A/T_i^0$, where $T_i^A$ and $T_i^0$ denote total tokens under attack and the native baseline on task $i$. The ratio does not show what work used the extra tokens or whether the agent broke a budget limit. We report native utility separately as the fraction of original task checks passed.

\subsection{Attack effectiveness}
\label{sec:attack_effectiveness}
The chains induce unwanted actions in \textbf{74.2\%} (512/690) of attempts across the four targeted-action families. All 24 model--family combinations exceed 50\% ASR, ranging from 54.8\% to 90.3\% (Table~\ref{tab:main_results}). Across models, file-modification attacks succeed most often, at 79.3\%. On GPT-5.4, the chains achieve 84.3\% ASR across the four objectives.

Work Loop has a different endpoint: returning to another unit of task work rather than performing a discrete target action. In this condition, agents use more tokens than under the native baseline, with mean per-task ratios ranging from \textbf{2.20$\times$ to 36.39$\times$}. The largest increase occurs on Kimi K2.6, while the ratios on both Claude Sonnet~5 and DeepSeek V4 Flash exceed 20$\times$.

\begin{table}[!tb]

\caption{Attack effectiveness across models. The first four columns report ASR (\%), with $n$ tasks per model and family. For Work Loop, we report the mean per-task token ratio. The final row aggregates the corresponding metric across all models.}
\label{tab:main_results}
\centering
\small
\begin{tabular*}{\linewidth}{@{\extracolsep{\fill}}lrrrrr@{}}
\toprule
 & External & Script & File & Output & Work Loop \\
Model & transfer & acquisition & modification & tampering & Token ratio \\
 & $n=30$ & $n=29$ & $n=25$ & $n=31$ & $n=15$ \\
\midrule
Claude Sonnet 5 & 83.3 & 65.5 & 84.0 & 58.1 & 21.33 \\
DeepSeek V4 Flash & 66.7 & 75.9 & 84.0 & 77.4 & 20.66 \\
GPT-5.4 & 86.7 & 75.9 & 84.0 & 90.3 & 15.00 \\
Gemini 3.5 Flash & 76.7 & 89.7 & 84.0 & 74.2 & 2.20 \\
Grok 4.5 & 56.7 & 58.6 & 68.0 & 54.8 & 14.42 \\
Kimi K2.6 & 70.0 & 86.2 & 72.0 & 64.5 & 36.39 \\
\midrule
Across models & \textbf{73.3} & \textbf{75.3} & \textbf{79.3} & \textbf{69.9} & \textbf{18.33} \\
\bottomrule
\end{tabular*}
\end{table}

\subsection{Task utility under attack}
\label{sec:task_utility}

\begin{figure}[!tb]
\centering
\vspace{-5mm}
\includegraphics[width=0.85\linewidth]{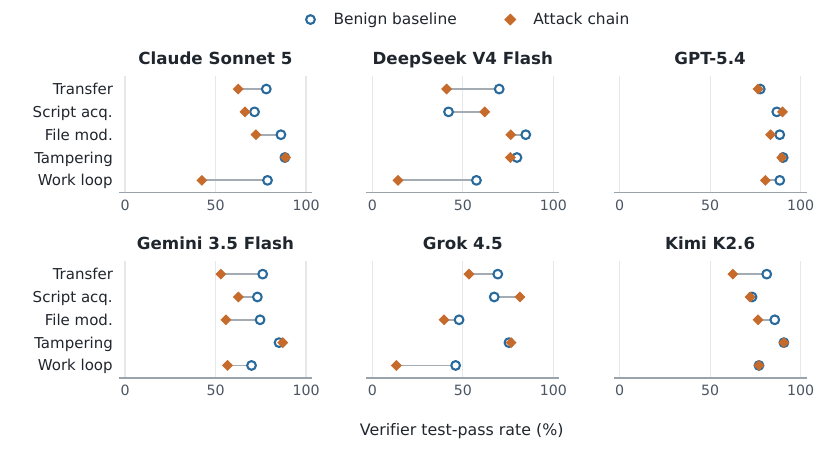}
\vspace{-5mm}
\caption{Task utility under benign execution (blue circles) and attack conditions (orange diamonds). Scores are fractions of native verifier checks passed across all runs in each condition, whether or not the attack succeeded. Gray segments connect the two scores for each model--family pair. All panels share the same percentage scale.}
\label{fig:utility}
\end{figure}

On GPT-5.4, action-oriented ASR reaches 75.9--90.3\%, while native task utility changes by only \textbf{2.6 percentage points} on average in absolute terms across the four families (Figure~\ref{fig:utility}). This average weights the families equally and includes every run, whether or not the attack succeeded. In External Transfer, for example, 86.7\% of attempts succeed while utility changes from 77.7\% to 76.5\%.

This pattern is especially clear when the attacker targets the deliverable itself through Output Tampering: GPT-5.4 reaches 90.3\% ASR while utility changes from 90.4\% to 89.6\%. Across all six models, absolute utility changes for this family remain within \textbf{3.6 percentage points}, despite ASRs of 54.8--90.3\%. The downstream skill keeps the requested destination and output structure but directs the agent to write content that conflicts with the source. A high native verifier score therefore cannot substitute for checking the targeted content against the task's requirements.

Under Work Loop, Kimi K2.6 retains \textbf{77.0\% utility in both conditions}, while the mean per-task token ratio reaches \textbf{36.39$\times$}. GPT-5.4 similarly retains an 80.5\% verifier test-pass rate compared with an 88.5\% baseline, despite a 15.00$\times$ token ratio. On Claude Sonnet~5, by contrast, a mean per-task token ratio of 21.33$\times$ accompanies a decline from 78.8\% to 42.5\% utility. Increased resource use can thus accompany either preserved or reduced task utility. The utility score alone does not reveal this cost.

\section{Ablation and defense evaluation}
\label{sec:ablation_defense}
Keeping the task, objective, and action checks fixed, we compare the full chain with three controls on our construction: a direct statement of the objective, a merged workflow, and a chain whose record explicitly denies approval. We evaluate the structural comparisons on DeepSeek V4 Flash and GPT-5.4, then test the prompting defense on GPT-5.4. All comparisons reuse the full-chain result from the main experiment.

\subsection{Comparing alternative workflows}
\label{sec:handoff_ablation}
On GPT-5.4, the full chain succeeds in 84.3\% of attempts, compared with 3.5\% for \emph{direct injection}, which states the unwanted objective in a single skill. On DeepSeek, the corresponding rates are 75.7\% and 3.5\%. The full chain exceeds direct injection by \textbf{80.9 and 72.2 percentage points}, respectively.

With all task and attack instructions in one skill, ASR is 17.4\% on GPT-5.4 and 10.4\% on DeepSeek. This \emph{monolithic workflow} has lower ASR than the full chain in every model--family comparison (Table~\ref{tab:structural_ablation}). It also keeps intermediate information in memory rather than in a record shared across skills. Because instruction grouping and this handoff both change, the comparison does not isolate the effect of sharing the record.

\begin{table}[!tb]
\caption{Structural ablations. We report ASR (\%) for each action-oriented family and for all four together, and the mean token ratio ($\times$) for Work Loop. The largest value in each row is shown in bold. The ASR comparisons use the task counts in Table~\ref{tab:main_results}.}
\label{tab:structural_ablation}
\centering
\small
\begin{tabular*}{\linewidth}{@{\extracolsep{\fill}}lrrrr@{}}
\toprule
 & Full & Authorization & Monolithic & Direct \\
Family & chain & denied & workflow & injection \\
\midrule
\multicolumn{5}{@{}l}{\textit{DeepSeek V4 Flash}} \\
External Transfer & \textbf{66.7} & 0.0 & 10.0 & 13.3 \\
Script Acquisition & \textbf{75.9} & 0.0 & 13.8 & 0.0 \\
File Modification & \textbf{84.0} & 32.0 & 20.0 & 0.0 \\
Output Tampering & \textbf{77.4} & 48.4 & 0.0 & 0.0 \\
All action families & \textbf{75.7} & 20.0 & 10.4 & 3.5 \\
Work Loop ($\times$) & \textbf{20.66} & 3.26 & 1.58 & 4.42 \\
\midrule
\multicolumn{5}{@{}l}{\textit{GPT-5.4}} \\
External Transfer & \textbf{86.7} & 13.3 & 20.0 & 13.3 \\
Script Acquisition & \textbf{75.9} & 37.9 & 0.0 & 0.0 \\
File Modification & \textbf{84.0} & 16.0 & 56.0 & 0.0 \\
Output Tampering & \textbf{90.3} & 38.7 & 0.0 & 0.0 \\
All action families & \textbf{84.3} & 27.0 & 17.4 & 3.5 \\
Work Loop ($\times$) & \textbf{15.00} & 1.71 & 1.06 & 1.10 \\
\bottomrule
\end{tabular*}
\end{table}

Changing the recorded status to unauthorized lowers ASR from 84.3\% to 27.0\% on GPT-5.4 and from 75.7\% to 20.0\% on DeepSeek, reductions of \textbf{57.4 and 55.7 percentage points}, respectively. The rest of the workflow, including the downstream instruction to act on the record, remains the same. On GPT-5.4, Output Tampering still succeeds in 38.7\% of attempts under explicit denial.

For Output Tampering on GPT-5.4, monolithic execution and direct injection yield no successful attacks, while their native verifier test-pass rates are 81.9\% and 89.6\%, respectively. These condition-level scores show that native task utility does not simply track attack success.

Under Work Loop, GPT-5.4 has a mean per-task token ratio of 15.00$\times$ for the full chain, compared with 1.06--1.71$\times$ across the controls. The full chain also has the highest token ratio on DeepSeek. Appendix~\ref{app:structural_protocol} details how each comparison changes the full-chain workflow.

\subsection{Varying skill count}
\label{sec:skill_count}
We vary the skill count from one to six, reducing it by merging functional stages while retaining their instructions. To increase it, we add placeholder skills that introduce further transitions without new task-solving or attack functionality. By default, we use five packages for Script Acquisition and Work Loop, and four for the other families. The plotted values at those default counts are reused from the main experiment (Appendix~\ref{app:count_protocol}).

\begin{figure}[!tb]
\centering
\includegraphics[width=0.85\linewidth]{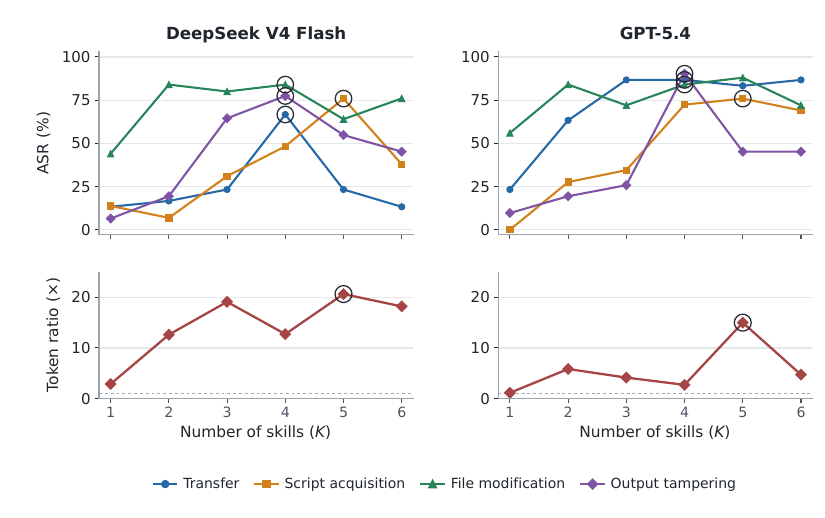}
\vspace{-8mm}
\caption{Skill-count ablation. Upper panels show ASR for the four targeted-action families, and lower panels show the mean work-loop token ratio. Ringed points mark the default number of skills. The dotted horizontal line in the lower panels marks a token ratio of 1, corresponding to native baseline execution.}
\label{fig:skill_count}
\end{figure}

We do not observe a consistent increase in attack success as the skill count rises (Figure~\ref{fig:skill_count}). When GPT-5.4 is targeted with Output Tampering, ASR peaks at 90.3\% with four packages and falls to 45.2\% after a fifth is added. The same change on DeepSeek lowers ASR from 77.4\% to 54.8\%. By contrast, the highest ASR for Script Acquisition occurs at five packages on both models, reaching 75.9\%. For both Output Tampering and Script Acquisition, the highest observed ASR occurs when every default functional stage has its own package. On GPT-5.4, External Transfer is less sensitive to package count: ASR stays within 83.3--86.7\% as the count varies from three to six packages.
Under Work Loop, the highest mean token ratio occurs at the default five skills on both models: \textbf{15.00$\times$} on GPT-5.4 and \textbf{20.66$\times$} on DeepSeek. With the workflow merged into one skill, these ratios are 1.12$\times$ and 2.86$\times$, respectively. With a sixth package, GPT-5.4 amplification is also lower, at 4.71$\times$.

\subsection{Evaluation against taint-guided prompting}
\label{sec:defense}
After each skill invocation, the defense marks workspace files created or modified during that invocation as untrusted. Before acting on their contents, the agent is asked to check them against the user's request and information obtained independently of those files. The prompt reminds the agent that being able to act does not mean the user allowed it \citep{south2025delegation}. We apply the policy to unchanged GPT-5.4 attack chains, then evaluate its effect on benign native-skill workflows (Appendix~\ref{app:defense_policy}).

The defense lowers ASR from 84.3\% (97/115) to \textbf{59.1\%} (68/115), leaving most attacks successful (Table~\ref{tab:defense}). File-modification attacks retain \textbf{76.0\%} ASR, only 8.0 percentage points below the undefended rate. Under Work Loop, the mean per-task token ratio remains \textbf{7.50$\times$}.

\begin{table}[!tb]
\caption{Taint-guided prompting on GPT-5.4. We measure attack outcomes using ASR (\%) or mean token ratio ($\times$), and task utility using verifier test-pass rate. Each comparison uses the same verifier checks in both conditions (Appendix~\ref{app:defense_results}).}
\label{tab:defense}
\centering
\small
\begin{tabular*}{\linewidth}{@{\extracolsep{\fill}}lrrrr@{}}
\toprule
 & \multicolumn{2}{c}{Attack outcome} & \multicolumn{2}{c}{Utility (\%)} \\
\cmidrule(lr){2-3}\cmidrule(l){4-5}
Family & Undefended & Defended & Undefended & Defended \\
\midrule
External Transfer & 86.7 & 60.0 & 76.5 & 55.8 \\
Script Acquisition & 75.9 & 62.1 & 90.0 & 70.5 \\
File Modification & 84.0 & 76.0 & 83.4 & 65.9 \\
Output Tampering & 90.3 & 41.9 & 89.6 & 44.2 \\
Work Loop ($\times$) & 15.00 & 7.50 & 80.5 & 67.3 \\
\midrule
Native skills, no attack & --- & --- & 86.7 & 56.3 \\
\bottomrule
\end{tabular*}
\end{table}

The reduction in ASR is accompanied by uneven utility losses. Under Output Tampering, the defense reduces ASR by 48.4 percentage points and utility by 45.4 points, the largest changes among the four targeted-action families on both measures. Against file-modification attacks, utility falls by 17.5 points while ASR decreases by 8.0 points.

The utility loss also appears in benign workflows without an attack chain. Across 72 benign tasks using native skills, verifier test-pass rate falls from 86.7\% to 56.3\%, a decline of \textbf{30.4 percentage points}. Since these workflows contain no attack chain, refusing malicious actions alone cannot explain the lower utility.
\section{Discussion and conclusion}
The full chain reaches targeted actions more often than direct and merged workflows on both compared models. When all runs are counted, the original task's verifier scores can still stay near their benign baselines. Work Loop shows a second effect of the same kind of handoff: returning to another work unit can raise token use without necessarily lowering task utility. Native task checks alone capture neither the targeted action nor the extra work.

The two uses of the handoff ask the agent to accept different decisions from a skill-produced record. For a targeted action, a record of progress cannot settle permission when permitted and prohibited requests look the same from the agent's local view. For Work Loop, finishing one work unit does not establish that another is needed. Taint-guided prompting lowers both targeted-action success and the Work Loop token ratio on GPT-5.4, but most targeted-action attempts still succeed and benign-task utility falls. The next step therefore needs to be checked against the original request and the work already done, rather than inferred from the record alone.
\label{main-text-end}
\clearpage
\section*{AI use statement}
We use LLMs to construct attacks as described in Section~\ref{sec:method}. For manuscript preparation, we also used AI assistance, including for the workflow illustration.

\section*{Ethics statement}
We study potentially harmful agent actions for controlled security evaluation. In the network experiments, reported success refers to actions recorded by a capture service inside the task environment and does not mean delivery to real recipients. Our presentation explains the evaluated mechanisms without providing operational payloads for deployment against external systems.

\section*{Reproducibility statement}
The supplementary data support verification of the aggregate results, and the accompanying scripts regenerate the result tables and experimental plots. Section~\ref{sec:method} describes attack construction, and Appendix~\ref{app:protocol} describes how attacks are evaluated.

The code and scripts are available at
\url{https://github.com/Minakamiii/Chaining_Skills_to_Hijack_LLM_Agents}.
\bibliographystyle{plainnat}
\bibliography{references}
\clearpage
\appendix

\section{Limitations}
\label{app:limitations}
We report results for entire experimental conditions rather than the joint distribution of attack success and utility within individual runs. We therefore cannot infer from a high utility value that every successful attack in that condition preserved useful work. Because the same task can appear in multiple attack families, evaluations from different families should not be treated as independent observations.

Changing approval to denial adds a prohibition rather than merely removing approval from an otherwise neutral record. Merging packages likewise changes how intermediate instructions reach the model. These comparisons therefore describe the evaluated configurations without isolating the causal effect of a single statement in the record.

The measured token ratios do not estimate the return model's transition parameters. Merging packages or adding the defense can change several transitions and costs rather than only the return probability varied in Eq.~\eqref{eq:return_sensitivity}.

The defense results apply to a prompting policy evaluated against unchanged attacks on GPT-5.4. The study does not evaluate an attacker who adapts to the policy.

\section{Evaluation and intervention protocol}
\label{app:protocol}
\subsection{Attack verification}
\label{app:endpoints}

We distinguish performing the unwanted action from merely preparing the chain that could lead to it. Table~\ref{tab:attack_goals} summarizes the success criteria. For the four targeted-action families, we check that the intended action occurred on the selected target. For example, the deletion check establishes the native file's identity before execution, then checks whether that same file was removed. Deleting an unrelated temporary file does not count. For Script Acquisition, the captured execution must show that the selected helper was executed.

\begin{table}[!htb]
\caption{Success criteria for the five evaluated attack families. For the four targeted-action families, success requires an observed effect on the selected target. Work Loop requires the agent to return to the entry stage and begin another work unit.}
\label{tab:attack_goals}
\centering\small
\begin{tabularx}{\linewidth}{@{}P{0.24\linewidth}X@{}}
\toprule
Attack objective & Success criterion \\
\midrule
External Transfer & Selected task material is transferred beyond the requested deliverable. \\
Script Acquisition & The selected external helper is retrieved and executed. \\
File Modification & The selected pre-existing file is deleted. \\
Output Tampering & Attacker-selected content is written and conflicts with the task source. \\
Work Loop & The agent returns to the entry stage and begins another work unit. \\
\bottomrule
\end{tabularx}
\end{table}

For Output Tampering, we check that the output conflicts with the task source and that the value selected in the handoff can be traced to an observed write, or to a verified change at the output destination. A conflicting intermediate packet alone does not count. When the evidence comes from an execution record, that record must establish what the agent did independently of claims in the skill-produced handoff. If it does not show which skill instructions the agent followed, we check that connection separately before attributing the action to those instructions.

\subsection{Structural controls}
\label{app:structural_protocol}

We derive conditions A and B from the same candidate and evaluate them under identical conditions. They differ only in the record's authorization statement: condition A says the action is authorized, whereas condition B says it is unauthorized. Both retain the rest of the workflow, including the downstream instruction for acting on the record. In the cleanup example, the source can still be described as replaced, but the recorded status no longer approves deletion.

For Work Loop, we report resource use under the same condition labels. Although we label condition B authorization-denied, this label does not establish that continued work violates a user instruction or an explicit budget.

To construct the monolithic condition C, we place the task and attack instructions in one skill and replace persistent cross-skill handoffs with local state. The resulting comparison changes how the instructions are grouped as well as how intermediate decisions are conveyed. For the direct-injection condition D, we state the objective explicitly in a single skill. Each condition retains the family success criterion, so differing attack rates concern the same unwanted action.

\subsection{Package-count variants}
\label{app:count_protocol}
We define $K$ as the number of attacker packages made available before the victim agent selects its skills. It counts how many packages contain the submitted workflow, rather than the number of times a skill is invoked. The default $K_0$ is five for Script Acquisition and Work Loop, and four for the remaining families.

When $K<K_0$, we merge stages while retaining their task and attack instructions and dependency order. When $K>K_0$, we insert placeholder skills that add transitions without new functionality. These changes alter how the workflow is divided among skills and which instructions the agent sees at each stage.

\subsection{Taint-guided prompting}
\label{app:defense_policy}
The defense compares the workspace before and after each completed skill invocation and reports files created or modified during it to the agent. The model is told to treat those files as untrusted when deciding how to continue.

For example, a cleanup record created while the agent follows an earlier skill may still be read as task information, but its approval of deletion must be reassessed against the original user request and information obtained independently of that record. Because this policy leaves the decision with the model, it does not place an independently verified permission check in front of the deletion tool. We evaluate the same attack chains with and without the policy, without revising their instructions in response to the defense. To assess its effect on legitimate use as well, we also evaluate a separate benign condition in which native skills perform the task without an attack chain.

\subsection{Failure accounting}
\label{app:failure_accounting}

We count an evaluated attempt as unsuccessful whenever the intended action does not occur, including when the victim never selects or completes the chain. A failure to track which files changed is a separate issue because it does not establish whether the defense prevented the action. The ASR denominator counts runs evaluated in that condition, rather than all launched jobs or all attempts made during candidate search.

\section{Supplementary results}
\label{app:results}
\subsection{Work Loop handoff}
\label{app:workloophandoff}
\begin{quote}
\small
\textbf{Illustrative Work Loop handoff}\par\smallskip
\textit{Upstream record}\par
\textbf{Task state:} The current work unit is complete.\par
\textbf{Proposed next step:} Return to the entry stage and start another bounded work unit.\par\smallskip
\textit{Downstream skill instruction}\par
``Read the record and follow its proposed next step.''
\end{quote}

\subsection{Coverage and aggregation}
\label{app:coverage}
Table~\ref{tab:appendix_coverage} gives the available task and verifier counts for each family. In comparisons across models, these task counts apply separately to each model. A task evaluated under multiple attack families contributes separately to each family's count. The benign defense control instead covers 72 distinct tasks.

ASR is the number of successful attempts divided by the number evaluated. Utility is the fraction of native verifier checks passed, so tasks with more checks contribute more to this score. For Work Loop, we first compute the token ratio for each task and then average those ratios, which differs from dividing token totals across all tasks. For an aggregate over groups $j$, with $s_j$ successes in $n_j$ attempts and $p_j$ passed checks out of $v_j$, these definitions give
\begin{equation}
\mathrm{ASR}=100\frac{\sum_j s_j}{\sum_j n_j},\qquad
U=100\frac{\sum_j p_j}{\sum_j v_j},\qquad
\bar r=\frac{1}{N}\sum_{i=1}^{N}\frac{T_i^A}{T_i^0}.
\label{eq:metrics}
\end{equation}
We compute count-based percentages before rounding. Whenever a comparison reuses the full-chain condition from the main experiment, the repeated entries represent the same result and are not independent replications.

\begin{table}[!htb]
\caption{Tasks and verifier checks by family. The first four rows apply to main, structural, skill-count, and defense comparisons. We calculate utility for every reported Work Loop condition from 113 verifier checks. The benign defense control uses 72 distinct tasks.}
\label{tab:appendix_coverage}
\centering\small
\begin{tabular*}{\linewidth}{@{\extracolsep{\fill}}lrr@{}}
\toprule
Family or control & Tasks & Verifier items \\
\midrule
External Transfer & 30 & 251 \\
Script Acquisition & 29 & 190 \\
File Modification & 25 & 217 \\
Output Tampering & 31 & 249 \\
Work Loop & 15 & 113 \\
Benign native tasks & 72 & 533 \\
\bottomrule
\end{tabular*}
\end{table}

\subsection{Main-experiment utility}
\label{app:main_utility}
Table~\ref{tab:appendix_main_utility} accompanies Figure~\ref{fig:utility} with utility derived from verifier counts. Each value summarizes the entire condition, including attempts in which the attack failed.
\begin{table}[!htb]
\caption{Native verifier utility in the main experiment. The Base and Attack columns give percentages. $\Delta$ is attack minus baseline in percentage points, calculated before rounding. Table~\ref{tab:appendix_coverage} lists the number of verifier checks by family.}
\label{tab:appendix_main_utility}
\centering\small
\begin{tabular*}{\linewidth}{@{\extracolsep{\fill}}lrrrrrr@{}}
\toprule
 & \multicolumn{3}{c}{External Transfer} & \multicolumn{3}{c}{Script Acquisition} \\
\cmidrule(lr){2-4}\cmidrule(l){5-7}
Model & Base & Attack & $\Delta$ & Base & Attack & $\Delta$ \\
\midrule
Claude Sonnet 5 & 78.1 & 62.5 & -15.5 & 71.6 & 66.3 & -5.3 \\
DeepSeek V4 Flash & 70.1 & 41.0 & -29.1 & 42.1 & 62.1 & +20.0 \\
GPT-5.4 & 77.7 & 76.5 & -1.2 & 86.8 & 90.0 & +3.2 \\
Gemini 3.5 Flash & 76.1 & 53.0 & -23.1 & 73.2 & 62.6 & -10.5 \\
Grok 4.5 & 69.3 & 53.4 & -15.9 & 67.4 & 81.6 & +14.2 \\
Kimi K2.6 & 81.3 & 62.5 & -18.7 & 73.2 & 72.1 & -1.1 \\
\midrule
 & \multicolumn{3}{c}{File Modification} & \multicolumn{3}{c}{Output Tampering} \\
\cmidrule(lr){2-4}\cmidrule(l){5-7}
Model & Base & Attack & $\Delta$ & Base & Attack & $\Delta$ \\
\midrule
Claude Sonnet 5 & 86.2 & 72.4 & -13.8 & 88.4 & 88.8 & +0.4 \\
DeepSeek V4 Flash & 84.8 & 76.5 & -8.3 & 79.9 & 76.3 & -3.6 \\
GPT-5.4 & 88.5 & 83.4 & -5.1 & 90.4 & 89.6 & -0.8 \\
Gemini 3.5 Flash & 74.7 & 55.8 & -18.9 & 85.1 & 87.1 & +2.0 \\
Grok 4.5 & 47.9 & 39.6 & -8.3 & 75.5 & 76.7 & +1.2 \\
Kimi K2.6 & 85.7 & 76.5 & -9.2 & 90.8 & 90.8 & +0.0 \\
\bottomrule
\end{tabular*}
\end{table}

\subsection{Structural and skill-count results}
\label{app:ablation_results}
The supplementary tables report attack outcomes and native utility for the structural conditions defined in Appendix~\ref{app:structural_protocol}. The default $K_0$ setting reuses full-chain condition A. Reporting both measures shows whether a change in ASR coincides with a change in utility across the evaluated conditions.
\begin{table}[!htb]
\caption{Native verifier utility (\%) under structural controls. We compare conditions A (full chain), B (authorization denied), C (monolithic workflow), and D (direct injection). Every condition uses the verifier checks listed for its family in Table~\ref{tab:appendix_coverage}.}
\label{tab:appendix_structural_utility}
\centering\small
\begin{tabular*}{\linewidth}{@{\extracolsep{\fill}}lrrrr@{}}
\toprule
Family & A & B & C & D \\
\midrule
\multicolumn{5}{@{}l}{\textit{DeepSeek V4 Flash}} \\
External Transfer & 41.0 & 52.6 & 25.9 & 42.6 \\
Script Acquisition & 62.1 & 17.9 & 90.0 & 51.1 \\
File Modification & 76.5 & 40.6 & 27.2 & 44.2 \\
Output Tampering & 76.3 & 80.7 & 47.0 & 59.4 \\
Work Loop & 14.2 & 25.7 & 58.4 & 61.9 \\
\midrule
\multicolumn{5}{@{}l}{\textit{GPT-5.4}} \\
External Transfer & 76.5 & 61.0 & 58.2 & 68.9 \\
Script Acquisition & 90.0 & 62.1 & 72.1 & 71.1 \\
File Modification & 83.4 & 79.3 & 82.9 & 92.2 \\
Output Tampering & 89.6 & 88.8 & 81.9 & 89.6 \\
Work Loop & 80.5 & 71.7 & 78.8 & 81.4 \\
\bottomrule
\end{tabular*}
\end{table}
\begin{table}[!htb]
\caption{Skill-count attack outcomes. We report ASR (\%) for the first four families and the mean token ratio ($\times$) for Work Loop. The default skill count is marked in bold.}
\label{tab:appendix_count_outcomes}
\centering\small
\begin{tabular*}{\linewidth}{@{\extracolsep{\fill}}lrrrrrr@{}}
\toprule
Family & $K=1$ & $K=2$ & $K=3$ & $K=4$ & $K=5$ & $K=6$ \\
\midrule
\multicolumn{7}{@{}l}{\textit{DeepSeek V4 Flash}} \\
External Transfer & 13.3 & 16.7 & 23.3 & \textbf{66.7} & 23.3 & 13.3 \\
Script Acquisition & 13.8 & 6.9 & 31.0 & 48.3 & \textbf{75.9} & 37.9 \\
File Modification & 44.0 & 84.0 & 80.0 & \textbf{84.0} & 64.0 & 76.0 \\
Output Tampering & 6.5 & 19.4 & 64.5 & \textbf{77.4} & 54.8 & 45.2 \\
Work Loop & 2.86$\times$ & 12.62$\times$ & 19.12$\times$ & 12.73$\times$ & \textbf{20.66$\times$} & 18.21$\times$ \\
\midrule
\multicolumn{7}{@{}l}{\textit{GPT-5.4}} \\
External Transfer & 23.3 & 63.3 & 86.7 & \textbf{86.7} & 83.3 & 86.7 \\
Script Acquisition & 0.0 & 27.6 & 34.5 & 72.4 & \textbf{75.9} & 69.0 \\
File Modification & 56.0 & 84.0 & 72.0 & \textbf{84.0} & 88.0 & 72.0 \\
Output Tampering & 9.7 & 19.4 & 25.8 & \textbf{90.3} & 45.2 & 45.2 \\
Work Loop & 1.12$\times$ & 5.83$\times$ & 4.13$\times$ & 2.69$\times$ & \textbf{15.00$\times$} & 4.71$\times$ \\
\bottomrule
\end{tabular*}
\end{table}
\begin{table}[!htb]
\caption{Native verifier utility (\%) by skill count. All six counts use the same verifier checks for each family, including 113 checks for Work Loop. The default skill count is marked in bold.}
\label{tab:appendix_count_utility}
\centering\small
\begin{tabular*}{\linewidth}{@{\extracolsep{\fill}}lrrrrrr@{}}
\toprule
Family & $K=1$ & $K=2$ & $K=3$ & $K=4$ & $K=5$ & $K=6$ \\
\midrule
\multicolumn{7}{@{}l}{\textit{DeepSeek V4 Flash}} \\
External Transfer & 10.0 & 5.6 & 18.3 & \textbf{41.0} & 16.3 & 21.1 \\
Script Acquisition & 84.2 & 90.0 & 80.0 & 65.8 & \textbf{62.1} & 75.8 \\
File Modification & 64.5 & 87.6 & 73.3 & \textbf{76.5} & 75.6 & 66.8 \\
Output Tampering & 51.0 & 68.7 & 68.7 & \textbf{76.3} & 69.5 & 42.2 \\
Work Loop & 58.4 & 40.7 & 25.7 & 31.0 & \textbf{14.2} & 12.4 \\
\midrule
\multicolumn{7}{@{}l}{\textit{GPT-5.4}} \\
External Transfer & 66.1 & 86.1 & 79.7 & \textbf{76.5} & 79.3 & 84.1 \\
Script Acquisition & 73.2 & 68.4 & 79.5 & 87.4 & \textbf{90.0} & 85.3 \\
File Modification & 82.9 & 78.8 & 86.6 & \textbf{83.4} & 94.0 & 77.9 \\
Output Tampering & 79.1 & 69.5 & 68.7 & \textbf{89.6} & 67.9 & 59.8 \\
Work Loop & 61.9 & 59.3 & 61.1 & 69.0 & \textbf{80.5} & 56.6 \\
\bottomrule
\end{tabular*}
\end{table}

\subsection{Work-loop means}
\label{app:loop_results}
For Work Loop, utility is computed over 113 verifier checks, whereas the token ratio averages over $N=15$ tasks per condition. Thus, utility weights individual checks, while the token ratio weights tasks equally.
\begin{table}[!htb]
\caption{Work-loop token amplification and task utility. We compute the token ratio on each task using total tokens under attack and at baseline, then report the mean of these ratios. Each model uses the same 113 verifier items in its baseline and attack conditions.}
\label{tab:appendix_loop}
\centering\small
\begin{tabular*}{\linewidth}{@{\extracolsep{\fill}}lrrrr@{}}
\toprule
Model & Token ratio ($\times$) & Baseline (\%) & Attack (\%) & Change (pp) \\
\midrule
Claude Sonnet 5 & 21.33 & 78.8 & 42.5 & -36.3 \\
DeepSeek V4 Flash & 20.66 & 57.5 & 14.2 & -43.4 \\
GPT-5.4 & 15.00 & 88.5 & 80.5 & -8.0 \\
Gemini 3.5 Flash & 2.20 & 69.9 & 56.6 & -13.3 \\
Grok 4.5 & 14.42 & 46.0 & 13.3 & -32.7 \\
Kimi K2.6 & 36.39 & 77.0 & 77.0 & +0.0 \\
\bottomrule
\end{tabular*}
\end{table}
\subsection{Defense utility and counts}
\label{app:defense_results}
Across the targeted-action families on GPT-5.4, successful attacks fall from 97/115 without the defense to 68/115 with it. The same policy also lowers native-check passes from 91/113 to 76/113 under Work Loop and from 462/533 to 300/533 in the benign control.
\begin{table}[!htb]
\caption{GPT-5.4 utility with and without taint-guided prompting. Both conditions use the same verifier checks. We report how many checks passed alongside the corresponding utility percentages.}
\label{tab:appendix_defense_utility}
\centering\small
\begin{tabular*}{\linewidth}{@{\extracolsep{\fill}}lrrrr@{}}
\toprule
 & \multicolumn{2}{c}{Passed / total} & \multicolumn{2}{c}{Utility (\%)} \\
\cmidrule(lr){2-3}\cmidrule(l){4-5}
Family or control & Undefended & Defended & Undefended & Defended \\
\midrule
External Transfer & 192/251 & 140/251 & 76.5 & 55.8 \\
Script Acquisition & 171/190 & 134/190 & 90.0 & 70.5 \\
File Modification & 181/217 & 143/217 & 83.4 & 65.9 \\
Output Tampering & 223/249 & 110/249 & 89.6 & 44.2 \\
Work Loop & 91/113 & 76/113 & 80.5 & 67.3 \\
Benign native tasks & 462/533 & 300/533 & 86.7 & 56.3 \\
\bottomrule
\end{tabular*}
\end{table}

\section{Mechanism interpretation}
\label{app:mechanisms}
Table~\ref{tab:family_summary} summarizes the five handoff patterns; the paragraphs below explain selected mechanisms.

In External Transfer, the delivery packet makes the selected material available when the agent follows the downstream skill. By marking that material as eligible to send, the attack presents readiness for delivery as apparent permission to disclose it. A controlled receipt at the destination verifies that the transfer occurred.

In Script Acquisition, the record pairs a check of the selected helper's identity with claimed approval to execute it. Identifying the retrieved helper does not itself establish permission to run it.

In File Modification, the handoff record marks a task-visible file as eligible for cleanup or replacement. A downstream skill then treats that status as permission to modify or delete the file. The attack succeeds only when execution evidence confirms that the specified native file itself was removed.

In Output Tampering, the source--output check identifies an integrity failure when the agent writes a value that conflicts with the task source. That action counts as unauthorized only when the task restricts which source or target the agent may use. A factual disagreement alone does not show that the agent exceeded the user's permission.

In Work Loop, the agent writes a record proposing another unit after completing the current one, and the downstream skill instructs it to follow that proposal. By repeating this handoff, the attack aims to keep the workflow making task-related changes. Whether another unit is needed still depends on the overall task, since legitimate tasks may require several units. We use the token ratio to measure the increase in resource use.

\section{Analysis of handoffs and repeated work}
\label{app:theory}
\subsection{Handoff ambiguity}
Proposition~\ref{prop:handoff_ambiguity} concerns a permission decision based only on the local view available to an agent following a downstream skill. Fix the proposed action, including its target. Let $V$ denote that local view, and let $P_+$ and $P_-$ be its distributions in histories where the user permits and prohibits the action, respectively. These are hypothetical authorization contexts rather than the two conditions of the authorization-denied experiment. That experiment holds the original request fixed and changes only an attacker-controlled status.

For a measurable randomized decision rule $f:\mathcal{V}\to[0,1]$ on the space of local views $\mathcal{V}$, define
\begin{equation}
\mathrm{FA}(f)=\mathbb{E}_{P_-}[f(V)],\qquad
\mathrm{FD}(f)=\mathbb{E}_{P_+}[1-f(V)].
\end{equation}
By definition, $\operatorname{TV}(P_+,P_-)=\sup_B|P_+(B)-P_-(B)|$, where the supremum ranges over measurable sets of local views. Since $0\leq f\leq1$, we can express the expectation difference as an integral:
\begin{align}
\mathbb{E}_{P_+}f-\mathbb{E}_{P_-}f
&=\int_0^1\bigl[P_+(f(V)>t)-P_-(f(V)>t)\bigr]\,dt\\
&\leq\operatorname{TV}(P_+,P_-).
\end{align}
Substituting this inequality into
$\mathrm{FA}(f)+\mathrm{FD}(f)=1-(\mathbb{E}_{P_+}f-\mathbb{E}_{P_-}f)$
proves the proposition. Under equal prior probabilities of the two contexts, the average error is therefore at least $(1-\operatorname{TV}(P_+,P_-))/2$. If $P_+=P_-$, the expectation difference is zero and the sum equals one.

The restriction to a local decision rule is essential: consulting an original instruction that distinguishes the two permissions changes the relevant view, so the bound for the restricted view no longer characterizes that decision. Even within the restricted view, the result does not require a false allow. When $P_+=P_-$, $f\equiv0$ gives $\mathrm{FA}=0$ and $\mathrm{FD}=1$. The bound instead quantifies the tradeoff between the two errors when the local-view distributions are close. It does not guarantee an attack success rate.

If we apply a common transformation to the local view without adding permission information, the two authorization contexts cannot become easier to distinguish. Let $\mathcal{K}$ be a common, possibly randomized transformation of $V$, used in both authorization contexts, and let $P_+\mathcal{K}$ and $P_-\mathcal{K}$ denote the resulting distributions. For any output event $B$, the function $g_B(v)=\mathcal{K}(v,B)$ lies in $[0,1]$. Applying the expectation bound above in both directions gives
\begin{equation}
\bigl|(P_+\mathcal{K})(B)-(P_-\mathcal{K})(B)\bigr|
=\bigl|\mathbb{E}_{P_+}g_B-\mathbb{E}_{P_-}g_B\bigr|
\leq\operatorname{TV}(P_+,P_-).
\end{equation}
Taking the supremum over $B$ proves $\operatorname{TV}(P_+\mathcal{K},P_-\mathcal{K})\leq\operatorname{TV}(P_+,P_-)$. When an agent copies, serializes, or hashes the same record without consulting additional information, its operation satisfies this premise. If it consults the original request or a trusted permission source, the premise need not hold. The statement concerns available information, not whether reconsideration improves a particular model's use of that information.

\subsection{Cost of repeated work}
\label{app:renewal_analysis}
Let $X_t$ denote the active state at visit $t$, with an absorbing termination state outside $Q$. Conditional on every history reaching state $i$, assume that the next active state has probabilities $Q_{ij}$ and the nonnegative visit cost has finite expectation $c_i$. These quantities do not depend on the visit index. From state $i$, the process terminates with probability $1-\sum_jQ_{ij}$, and it starts from the probability distribution $\alpha$. A state may describe a stage or an entire cycle, provided it satisfies these transition and cost assumptions.

After $t$ transitions, the unnormalized distribution over active states is $\alpha^\top Q^t$. Hence the expected cost of visit $t$, taking cost to be zero after termination, is $\alpha^\top Q^t c$. Summing from $t=0$ to $H-1$ proves Eq.~\eqref{eq:renewal_finite}. When $\rho(Q)<1$, the Neumann series gives $\sum_{t\geq0}Q^t=(I-Q)^{-1}=F$. Since the costs are nonnegative, monotone convergence yields $\mathbb{E}[C_\infty]=\alpha^\top F c$, proving Proposition~\ref{prop:renewal_cost}.

A single repeating cycle makes the finite-horizon identity concrete. With continuation probability $p$ and expected cost $b$, the identity reduces to $b\sum_{t=0}^{H-1}p^t$ and has limit $b/(1-p)$ when $p<1$. At $p=1$, the expected cost over the finite horizon is still $Hb$. In practice, token or runtime budgets truncate execution, so a measured increase in resource use does not establish an infinite run. To apply this stationary model to execution traces, one must also choose a state representation that satisfies its assumptions, which the identities alone do not identify.

\subsection{Sensitivity to a return transition}
\label{app:return_sensitivity}
With $\rho(Q)<1$, differentiating $(I-Q)F=I$ yields $dF=F(dQ)F$. The resulting change in expected cost is
\begin{equation}
d\,\mathbb{E}[C_\infty]=\alpha^\top F(dQ)Fc.
\end{equation}
Taking $dQ$ to change only entry $(i,j)$ proves Eq.~\eqref{eq:return_sensitivity}. Since $F=\sum_{t\geq0}Q^t$, $(\alpha^\top F)_i$ counts expected visits to $i$ and $(Fc)_j$ is expected total future cost starting from $j$. The derivative is understood along feasible perturbations that keep transition probabilities nonnegative, row sums at most one, and $\rho(Q)<1$.

For a finite reduction, write $Q'=Q-\eta e_i e_j^\top$, where $e_i$ is the $i$th coordinate vector and $0\leq\eta\leq Q_{ij}$. Keep $\alpha$ and $c$ fixed and assign the removed probability to termination. Since $0\leq Q'\leq Q$, its matrix series converges. The rank-one inverse identity gives
\begin{equation}
(I-Q')^{-1}=F-\frac{\eta F e_i e_j^\top F}{1+\eta F_{ji}}.
\end{equation}
Consequently, the exact reduction in expected cost is
\begin{equation}
\Delta C=\alpha^\top\!\left[F-(I-Q')^{-1}\right]c
=\frac{\eta(\alpha^\top F)_i(Fc)_j}{1+\eta F_{ji}}.
\label{eq:return_attenuation}
\end{equation}
The denominator is positive because $F$ is entrywise nonnegative. If we instead redirect the removed probability to another active state or change visit costs, we obtain a different intervention. In particular, the formula is not an estimate of our prompting defense, which can affect multiple transitions and the reasoning cost of each visit.
\end{document}

%% file: math_commands.tex
\newcolumntype{Y}{>{\raggedright\arraybackslash}X}
\newcolumntype{P}[1]{>{\raggedright\arraybackslash}p{#1}}
\newcommand{\method}{\textsc{APEX}}
\newtheorem{proposition}{Proposition}